\documentclass[aps,prl,twocolumn,showpacs,floats]{revtex4}

\usepackage{epsfig}
\usepackage[dvips]{color}     
\usepackage[latin1]{inputenc} 
\usepackage{floatflt}

\newcommand{\tcb}{\textcolor{blue}} 
\newcommand{\tcr}{\textcolor{red}}

\newcommand{\be}{\begin{equation}}
\newcommand{\ee}{\end{equation}}
\newcommand{\bea}{\begin{eqnarray}}
\newcommand{\eea}{\end{eqnarray}}
\newcommand{\bt}{\begin{tabular}}
\newcommand{\et}{\end{tabular}}
\newcommand{\ba}{\begin{array}}
\newcommand{\ea}{\end{array}}
\newcommand{\ov}{\overline}
\newcommand{\wt}{\widetilde}
\newcommand{\bvec}{\mathbf}
\newcommand{\dy}{\displaystyle}
\newcommand{\vecna}{\mbox{\boldmath $\nabla$}}
\newcommand{\ellbf}{\mbox{\boldmath $\ell$}}

\begin{document}

\title{Superconductors with two critical temperatures}

\author{E. Di Grezia}
\email{digrezia@na.infn.it}
\author{S. Esposito}
\email{salvatore.esposito@na.infn.it}
\affiliation{Dipartimento di Scienze Fisiche dell'Universit\`a di
Napoli ``Federico II'' and Istituto Nazionale di Fisica Nucleare,
Sezione di Napoli, Complesso Universitario di Monte S. Angelo, Via
Cinthia, I-80126 Napoli, Italy}
\author{G. Salesi}
\email{salesi@unibg.it}
\affiliation{\mbox{Facolt\`a di Ingegneria dell'Universit\`a Statale di 
Bergamo, viale Marconi 5, I-24044 Dalmine (BG), Italy}\\
and Istituto Nazionale di Fisica Nucleare, Sezione di Milano, via
Celoria 16, I-20133 Milano, Italy}

\begin{abstract}
We propose a simple model for superconductors endowed with two critical
temperatures, corresponding to two second-order phase transitions,
in the framework of the Ginzburg-Landau mean-field theory.
For very large Cooper pair self-interaction, in addition to the standard
condensation occurring in the Ginzburg-Landau theory, we find another phase transition 
at a lower temperature with a maximum difference of 15\%
between the two critical temperatures. 


\end{abstract}

\maketitle

The Higgs mechanism \cite{Higgs,Bailin}, plays a basic role in the
macroscopic theory of gapped BCS superconductors, or in the
corresponding Ginzburg-Landau (GL) effective theory, where it
accounts for the emergence of short-range electromagnetic
interactions mediated by massive-like photons (responsible, for
example, of the Meissner effect) \cite{Tinkham}. The
main role in the Higgs mechanism is played by a scalar field, a
degree of freedom of which undergoes a ``condensation''. In this
letter we are going to see that the condensation of given degrees
of freedom rather than other ones is not a consequence of a mere
formal gauge choice, but can entail observable physical effects.

In GL theory, the dynamics of the Cooper pairs is ruled by a
complex order parameter $\phi$, which can be interpreted as the
wave function of the Cooper pair in its center-of-mass frame or,
at the same time, as the Higgs field for the electromagnetic U(1)
symmetry breaking. The lagrangian density describing the field
$\phi$ interacting with the electromagnetic field $A_\mu$ is given
by:
\begin{equation} {\cal L}=-\frac{1}{4}F_{\mu \nu }F^{\mu \nu
}+\left( D_{\mu }\phi \right)^{\dagger }\left( D^{\mu }\phi
\right) -U(\phi ,\phi^{\dagger })\,, \label{L}
\end{equation}
where \ $F_{\mu\nu}\equiv\partial_\mu A_\nu - \partial_\nu A_\mu$
accounts for the kinetic energy of the electromagnetic field,
\ and \ $D_{\mu }=\partial_{\mu }+2ieA_{\mu }$ \ is the covariant
derivative for the Cooper pair with electric charge $2e$. The
potential energy $U(\phi ,\phi^{\dagger })$ may be parameterized
as 
\be 
U(\phi ,\phi^{\dagger })=\lambda\,\left(|\phi|^{2}-|\phi_0|^{2}\right)^2
\ee 
in terms of a coupling constant $\lambda$ describing the self-interaction 
of the Cooper pairs and of a constant value $\phi_0$
corresponding to the minimum of $U$ ($m^2 = -2\lambda |\phi_0|^2$
is the mass parameter of the theory). 

Let us now study the small fluctuations of the scalar field
around the minimum of the potential energy and expand the complex $\phi$ as
follows:
\begin{equation}
\phi \equiv \frac{1}{\sqrt
2}(\eta_0+\eta)\,{\rm e}^{i\theta/\eta_0}\,, \label{6}
\end{equation}
where $\eta$ and $\theta$ are real fields. By inserting the 
parametrization (\ref{6}) in the above Lagrangian, a Bose condensation of the 
field $\eta$ takes place as a result of the spontaneous $U(1)$ symmetry
breaking. In this case, after the phase transition, the excitation of the 
field $\theta$ becomes directly the longitudinal degree of freedom of 
the massive gauge field $A_\mu$.
\ Actually, after performing the gauge transformation of the electromagnetic potential
\ $\displaystyle A_{\mu}\rightarrow A_{\mu }+e^{-1}\partial_{\mu }\theta$, \
the Lagrangian reads as follows: 
\begin{eqnarray}
&&{\cal L} \equiv {\cal L}_{A}+{\cal L}_{\eta }+{\cal L}_{h.o.} \label{11} \nonumber \\
&&{\cal L}_{A}=-\frac{1}{4}F_{\mu \nu }F^{\mu \nu }+4e^{2}\eta_0^2 A_\mu A^\mu 
\label{12} \nonumber\\
&&{\cal L}_{\eta }=(\partial_{\mu }\eta )(\partial^{\mu }\eta )-4\lambda
\eta_0^{2}\eta^{2} \nonumber\\
&&{\cal L}_{h.o.}=-4\lambda \eta_0\eta^{3}-\lambda \eta^{4}+8e^{2}
\eta_0\eta A_{\mu }A^{\mu }+4e^{2}\eta^{2}A_{\mu }A^{\mu }\,. \nonumber
\end{eqnarray} 
The lagrangian ${\cal L}_{A}$ is that typical of a photon with 
mass \ $m_{A}^{2}=8e^{2}\eta_0^{2}$, \ while ${\cal L}_\eta$ describes a
real scalar field $\eta$ with a mass \ $m_{\eta}^{2}=4\lambda\eta_0^{2}=-2m^{2}$. 
\ The phase field $\theta $ has disappeared from our Lagrangian due
to the previous gauge transformation.

From the above Lagrangian (it is sufficient to take only the 
quadratic terms in the fields) we can deduce the finite temperature effective 
potential and evaluate \cite{NXC} the critical temperature $T_1$ by requiring that the 
potential has a minimum for nonzero values of the $\phi$ VEV,
obtaining \footnote{Let us recall that an equal critical temperature has been obtained \cite{NXC}
also in the representation mostly used in the literature on Higgs mechanism:
the so-called  ``unitary gauge" \cite{Lee}.}: 
\begin{equation}
T_1^2 = \frac{4m^2}{\lambda + 4e^2}\,. \label{7}
\end{equation}
For $T>T_1$ the minimum of the effective potential is at
$\eta_0=0$ and no spontaneous symmetry breaking occurs. 
In this phase the condensate of the Cooper pairs can be described by a 
complex scalar field $\phi$ with a negative mass parameter $m^{2}$ (therefore the 
condensate shows a finite coherence lenght in the medium); while the 
infinite-range electromagnetic interactions are described by a massless photon 
field $A_{\mu}$.
Instead for $T<T_1$ the extremum at $\eta_0=0$ becomes a local maximum, while
a minimum for the effective potential arises at $\eta_0\not =0$,
and the passage of the system from one phase to the other at
$T=T_1$ is a second-order phase transition.
After the superconducting phase transition, one of the two degrees
of freedom of the complex scalar field $\phi$ ($\eta$ in our case) 
accounts for the fluctuations of the condensate of Cooper pairs, while the other
one (the phase $\theta$) has been ``eaten" by the gauge field $A_\mu$, 
become massive, and then describing finite-range electromagnetic interactions.\\
   
\noindent Taking a different representation of the scalar field $\phi$
obviously does not affect the number of degrees of freedom, but
can actually involve a slightly different broken symmetry
scenario. At variance with Eq.\,(\ref{6}), let us expand the
complex scalar field $\phi$ as:
\begin{equation}
\phi \equiv \frac{1}{\sqrt 2}(\phi_0 + \phi_{1} +i \phi_{2}),
\label{4}
\end{equation}
where $\phi_{1}, \phi_{2}$ are two real scalar fields,
and assume that the Bose condensation takes place in the field $\phi_1$ (or in $\phi_2$)
rather than in $\eta$.
Analogously to above we can insert the parametrization (\ref{4})
in Lagrangian (\ref{L}) obtaining this time the effective potential as a function 
of $\phi_0$, $\phi_1$, $\phi_2$: as a consequence the critical 
temperature $T_2$ for this phase transition, calculated (for a generic Higgs field) in 
\cite{NXC,Bailin}, results to be different from $T_1$
\begin{equation}
T_2^2 = \frac{3m^2}{\lambda + 3e^2} . \label{5}
\end{equation}
\noindent The reason for such a difference is made clear when expanding in
$\theta/\eta_0$ the exponential in Eq.\,(\ref{6}) and comparing
with Eq.\,(\ref{4}):
\begin{eqnarray}
\phi_0 & \sim & \eta_0
\nonumber \\
\phi_1 & \sim & \eta - \frac{\theta}{2}\left(\frac{\theta}{\eta_0}\right)+ \cdots
\label{8b} \\
\phi_2 & \sim & \theta \, + \, \eta\left(\frac{\theta}{\eta_0}\right)+ \cdots.
\nonumber
\end{eqnarray}
The degrees of freedom carried out by the real scalar fields
$\phi_1 ,\phi_2$ are different from those corresponding to $\eta
, \theta$, and tend to coincide only in the limit
$\eta_0\rightarrow \infty$. Actually, in Eqs.\,(\ref{8b}) the higher
orders in $\eta_0^{-1}$ contribute at the denominator of the expression
(\ref{7}) as an additional $\lambda/3$ term; that is, it arises an
increased effective self-interaction of the Cooper pairs ($\lambda\rightarrow
\lambda_{\rm eff}=4\lambda/3$).

Then we can see that, in general, two different superconducting
phases can arise in a given GL system, characterized by slightly
different critical temperatures $T_2, T_1$ pointing at two
second-order phase transitions. Such two phases correspond to
different condensations of electrons in Cooper pairs which exhibit
different self-interaction (ruled by the $\lambda$ coupling
constant), and described in the GL theory by
different scalar fields. The realization of one of the two regimes
is ruled by the relative strength of the Cooper pair
self-interaction with respect to the electromagnetic interaction.
By measuring the relative strength through the parameter $x=\lambda/4 e^2$, the ratio
of the two critical temperatures can be written as follows:
\begin{equation}
\frac{T_1}{T_2} = \sqrt{\frac{4x + 3}{3x + 3}}\geq 1\,, \label{9}
\end{equation}
the temperature $T_1$ [related to the representation in
(\ref{6})] being the largest one. When the Cooper pair
self-interaction is small compared to the electromagnetic interaction,
$\lambda \ll 4e^2$, the two critical temperatures almost coincide,
\begin{equation}
T_1 \simeq  \left(1 + \frac{x}{6}\right)T_2 , \label{10}
\end{equation}
and, practically, we observe only one effective phase transition.
On the other side, when $\lambda \gg 4e^2$, the ratio of the two
critical temperatures could be as large as $\sqrt{4/3} \simeq
1.15$,
\begin{equation}
\frac{T_1}{T_2} \simeq \sqrt{\frac{4}{3}} - \frac{1}{4 \sqrt 3
}\frac{1}{x}\,, \label{11}
\end{equation}
so that a $15\%$ effect may be expected in differentiating the
two superconducting phases. In Fig.\,1 we have plotted the ratio
$T_1/T_2$ as a function of $\lambda / 4 e^2$.

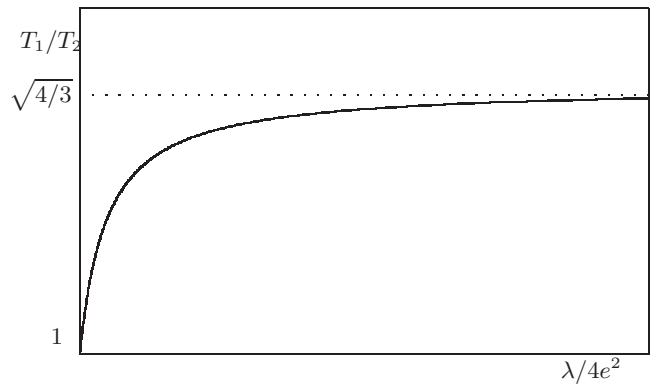
\begin{figure}
\begin{center}
\setlength{\unitlength}{0.240900pt}
\ifx\plotpoint\undefined\newsavebox{\plotpoint}\fi
\sbox{\plotpoint}{\rule[-0.200pt]{0.400pt}{0.400pt}}%
\begin{picture}(974,584)(0,0)
\font\gnuplot=cmr10 at 10pt
\gnuplot
\sbox{\plotpoint}{\rule[-0.200pt]{0.400pt}{0.400pt}}%
\put(60.0,41.0){\rule[-0.200pt]{0.400pt}{130.809pt}}
\put(60.0,41.0){\rule[-0.200pt]{0.400pt}{0.482pt}}
\put(60.0,582.0){\rule[-0.200pt]{0.400pt}{0.482pt}}
\put(60.0,41.0){\rule[-0.200pt]{215.124pt}{0.400pt}}
\put(953.0,41.0){\rule[-0.200pt]{0.400pt}{130.809pt}}
\put(60.0,584.0){\rule[-0.200pt]{215.124pt}{0.400pt}}
\put(-51,448){\makebox(0,0)[l]{$\sqrt{4/3}$}}
\put(15,68){\makebox(0,0)[l]{$1$}}
\put(864,14){\makebox(0,0){$\lambda/4e^2$}}
\put(-33,530){\makebox(0,0)[l]{$T_1/T_2$}}
\put(60.0,41.0){\rule[-0.200pt]{0.400pt}{130.809pt}}
\put(60,41){\usebox{\plotpoint}}
\put(60,41){\usebox{\plotpoint}}
\put(60,41){\usebox{\plotpoint}}
\put(60,41){\usebox{\plotpoint}}
\put(60,41){\usebox{\plotpoint}}
\put(60,41){\usebox{\plotpoint}}
\put(60.0,41.0){\rule[-0.200pt]{0.400pt}{1.204pt}}
\put(60.0,46.0){\usebox{\plotpoint}}
\put(61.0,46.0){\rule[-0.200pt]{0.400pt}{2.409pt}}
\put(61.0,56.0){\usebox{\plotpoint}}
\put(62.0,56.0){\rule[-0.200pt]{0.400pt}{2.168pt}}
\put(62.0,65.0){\usebox{\plotpoint}}
\put(63.0,65.0){\rule[-0.200pt]{0.400pt}{2.168pt}}
\put(63.0,74.0){\usebox{\plotpoint}}
\put(64.0,74.0){\rule[-0.200pt]{0.400pt}{1.927pt}}
\put(64.0,82.0){\usebox{\plotpoint}}
\put(65.0,82.0){\rule[-0.200pt]{0.400pt}{1.927pt}}
\put(65.0,90.0){\usebox{\plotpoint}}
\put(66.0,90.0){\rule[-0.200pt]{0.400pt}{1.927pt}}
\put(66.0,98.0){\usebox{\plotpoint}}
\put(67.0,98.0){\rule[-0.200pt]{0.400pt}{1.686pt}}
\put(67.0,105.0){\usebox{\plotpoint}}
\put(68.0,105.0){\rule[-0.200pt]{0.400pt}{1.686pt}}
\put(68.0,112.0){\usebox{\plotpoint}}
\put(69.0,112.0){\rule[-0.200pt]{0.400pt}{1.686pt}}
\put(69.0,119.0){\usebox{\plotpoint}}
\put(70.0,119.0){\rule[-0.200pt]{0.400pt}{1.686pt}}
\put(70.0,126.0){\usebox{\plotpoint}}
\put(71.0,126.0){\rule[-0.200pt]{0.400pt}{1.445pt}}
\put(71.0,132.0){\usebox{\plotpoint}}
\put(72.0,132.0){\rule[-0.200pt]{0.400pt}{1.445pt}}
\put(72.0,138.0){\usebox{\plotpoint}}
\put(73.0,138.0){\rule[-0.200pt]{0.400pt}{1.445pt}}
\put(73.0,144.0){\usebox{\plotpoint}}
\put(74.0,144.0){\rule[-0.200pt]{0.400pt}{1.445pt}}
\put(74.0,150.0){\usebox{\plotpoint}}
\put(75.0,150.0){\rule[-0.200pt]{0.400pt}{1.204pt}}
\put(75.0,155.0){\usebox{\plotpoint}}
\put(76.0,155.0){\rule[-0.200pt]{0.400pt}{1.204pt}}
\put(76.0,160.0){\usebox{\plotpoint}}
\put(77.0,160.0){\rule[-0.200pt]{0.400pt}{1.445pt}}
\put(77.0,166.0){\usebox{\plotpoint}}
\put(78.0,166.0){\rule[-0.200pt]{0.400pt}{0.964pt}}
\put(78.0,170.0){\usebox{\plotpoint}}
\put(79.0,170.0){\rule[-0.200pt]{0.400pt}{1.204pt}}
\put(79.0,175.0){\usebox{\plotpoint}}
\put(80.0,175.0){\rule[-0.200pt]{0.400pt}{1.204pt}}
\put(80.0,180.0){\usebox{\plotpoint}}
\put(81.0,180.0){\rule[-0.200pt]{0.400pt}{0.964pt}}
\put(81.0,184.0){\usebox{\plotpoint}}
\put(82.0,184.0){\rule[-0.200pt]{0.400pt}{1.204pt}}
\put(82.0,189.0){\usebox{\plotpoint}}
\put(83.0,189.0){\rule[-0.200pt]{0.400pt}{0.964pt}}
\put(83.0,193.0){\usebox{\plotpoint}}
\put(84.0,193.0){\rule[-0.200pt]{0.400pt}{0.964pt}}
\put(84.0,197.0){\usebox{\plotpoint}}
\put(85.0,197.0){\rule[-0.200pt]{0.400pt}{0.964pt}}
\put(85.0,201.0){\usebox{\plotpoint}}
\put(86.0,201.0){\rule[-0.200pt]{0.400pt}{0.964pt}}
\put(86.0,205.0){\usebox{\plotpoint}}
\put(87.0,205.0){\rule[-0.200pt]{0.400pt}{0.723pt}}
\put(87.0,208.0){\usebox{\plotpoint}}
\put(88.0,208.0){\rule[-0.200pt]{0.400pt}{0.964pt}}
\put(88.0,212.0){\usebox{\plotpoint}}
\put(89.0,212.0){\rule[-0.200pt]{0.400pt}{0.723pt}}
\put(89.0,215.0){\usebox{\plotpoint}}
\put(90.0,215.0){\rule[-0.200pt]{0.400pt}{0.964pt}}
\put(90.0,219.0){\usebox{\plotpoint}}
\put(91.0,219.0){\rule[-0.200pt]{0.400pt}{0.723pt}}
\put(91.0,222.0){\usebox{\plotpoint}}
\put(92.0,222.0){\rule[-0.200pt]{0.400pt}{0.723pt}}
\put(92.0,225.0){\usebox{\plotpoint}}
\put(93,227.67){\rule{0.241pt}{0.400pt}}
\multiput(93.00,227.17)(0.500,1.000){2}{\rule{0.120pt}{0.400pt}}
\put(93.0,225.0){\rule[-0.200pt]{0.400pt}{0.723pt}}
\put(94,229){\usebox{\plotpoint}}
\put(94,229){\usebox{\plotpoint}}
\put(94,229){\usebox{\plotpoint}}
\put(94,229){\usebox{\plotpoint}}
\put(94,229){\usebox{\plotpoint}}
\put(94,229){\usebox{\plotpoint}}
\put(94,229){\usebox{\plotpoint}}
\put(94,229){\usebox{\plotpoint}}
\put(94,229){\usebox{\plotpoint}}
\put(94,229){\usebox{\plotpoint}}
\put(94,229){\usebox{\plotpoint}}
\put(94,229){\usebox{\plotpoint}}
\put(94,229){\usebox{\plotpoint}}
\put(94,229){\usebox{\plotpoint}}
\put(94,229){\usebox{\plotpoint}}
\put(94,229){\usebox{\plotpoint}}
\put(94,229){\usebox{\plotpoint}}
\put(94,229){\usebox{\plotpoint}}
\put(94,229){\usebox{\plotpoint}}
\put(94,229){\usebox{\plotpoint}}
\put(94,229){\usebox{\plotpoint}}
\put(94,229){\usebox{\plotpoint}}
\put(94,229){\usebox{\plotpoint}}
\put(94,229){\usebox{\plotpoint}}
\put(94,229){\usebox{\plotpoint}}
\put(94,229){\usebox{\plotpoint}}
\put(94,229){\usebox{\plotpoint}}
\put(94,229){\usebox{\plotpoint}}
\put(94,229){\usebox{\plotpoint}}
\put(94,229){\usebox{\plotpoint}}
\put(94,229){\usebox{\plotpoint}}
\put(94,229){\usebox{\plotpoint}}
\put(94,229){\usebox{\plotpoint}}
\put(94,229){\usebox{\plotpoint}}
\put(94,229){\usebox{\plotpoint}}
\put(94.0,229.0){\rule[-0.200pt]{0.400pt}{0.723pt}}
\put(94.0,232.0){\usebox{\plotpoint}}
\put(95.0,232.0){\rule[-0.200pt]{0.400pt}{0.723pt}}
\put(95.0,235.0){\usebox{\plotpoint}}
\put(96.0,235.0){\rule[-0.200pt]{0.400pt}{0.482pt}}
\put(96.0,237.0){\usebox{\plotpoint}}
\put(97.0,237.0){\rule[-0.200pt]{0.400pt}{0.723pt}}
\put(97.0,240.0){\usebox{\plotpoint}}
\put(98.0,240.0){\rule[-0.200pt]{0.400pt}{0.723pt}}
\put(98.0,243.0){\usebox{\plotpoint}}
\put(99.0,243.0){\rule[-0.200pt]{0.400pt}{0.723pt}}
\put(99.0,246.0){\usebox{\plotpoint}}
\put(100.0,246.0){\rule[-0.200pt]{0.400pt}{0.482pt}}
\put(100.0,248.0){\usebox{\plotpoint}}
\put(101.0,248.0){\rule[-0.200pt]{0.400pt}{0.723pt}}
\put(101.0,251.0){\usebox{\plotpoint}}
\put(102.0,251.0){\rule[-0.200pt]{0.400pt}{0.482pt}}
\put(102.0,253.0){\usebox{\plotpoint}}
\put(103.0,253.0){\rule[-0.200pt]{0.400pt}{0.723pt}}
\put(103.0,256.0){\usebox{\plotpoint}}
\put(104.0,256.0){\rule[-0.200pt]{0.400pt}{0.482pt}}
\put(104.0,258.0){\usebox{\plotpoint}}
\put(105.0,258.0){\rule[-0.200pt]{0.400pt}{0.723pt}}
\put(105.0,261.0){\usebox{\plotpoint}}
\put(106.0,261.0){\rule[-0.200pt]{0.400pt}{0.482pt}}
\put(106.0,263.0){\usebox{\plotpoint}}
\put(107.0,263.0){\rule[-0.200pt]{0.400pt}{0.482pt}}
\put(107.0,265.0){\usebox{\plotpoint}}
\put(108.0,265.0){\rule[-0.200pt]{0.400pt}{0.482pt}}
\put(108.0,267.0){\usebox{\plotpoint}}
\put(109.0,267.0){\rule[-0.200pt]{0.400pt}{0.482pt}}
\put(109.0,269.0){\usebox{\plotpoint}}
\put(110.0,269.0){\rule[-0.200pt]{0.400pt}{0.482pt}}
\put(110.0,271.0){\usebox{\plotpoint}}
\put(111.0,271.0){\rule[-0.200pt]{0.400pt}{0.482pt}}
\put(111.0,273.0){\usebox{\plotpoint}}
\put(112.0,273.0){\rule[-0.200pt]{0.400pt}{0.482pt}}
\put(112.0,275.0){\usebox{\plotpoint}}
\put(113.0,275.0){\rule[-0.200pt]{0.400pt}{0.482pt}}
\put(113.0,277.0){\usebox{\plotpoint}}
\put(114.0,277.0){\rule[-0.200pt]{0.400pt}{0.482pt}}
\put(114.0,279.0){\usebox{\plotpoint}}
\put(115.0,279.0){\rule[-0.200pt]{0.400pt}{0.482pt}}
\put(115.0,281.0){\usebox{\plotpoint}}
\put(116.0,281.0){\rule[-0.200pt]{0.400pt}{0.482pt}}
\put(116.0,283.0){\usebox{\plotpoint}}
\put(117.0,283.0){\rule[-0.200pt]{0.400pt}{0.482pt}}
\put(117.0,285.0){\usebox{\plotpoint}}
\put(118.0,285.0){\rule[-0.200pt]{0.400pt}{0.482pt}}
\put(118.0,287.0){\usebox{\plotpoint}}
\put(119.0,287.0){\usebox{\plotpoint}}
\put(119.0,288.0){\usebox{\plotpoint}}
\put(120.0,288.0){\rule[-0.200pt]{0.400pt}{0.482pt}}
\put(120.0,290.0){\usebox{\plotpoint}}
\put(121.0,290.0){\rule[-0.200pt]{0.400pt}{0.482pt}}
\put(121.0,292.0){\usebox{\plotpoint}}
\put(122.0,292.0){\usebox{\plotpoint}}
\put(122.0,293.0){\usebox{\plotpoint}}
\put(123.0,293.0){\rule[-0.200pt]{0.400pt}{0.482pt}}
\put(123.0,295.0){\usebox{\plotpoint}}
\put(124.0,295.0){\usebox{\plotpoint}}
\put(124.0,296.0){\usebox{\plotpoint}}
\put(125.0,296.0){\rule[-0.200pt]{0.400pt}{0.482pt}}
\put(125.0,298.0){\usebox{\plotpoint}}
\put(126.0,298.0){\usebox{\plotpoint}}
\put(126.0,299.0){\usebox{\plotpoint}}
\put(127.0,299.0){\rule[-0.200pt]{0.400pt}{0.482pt}}
\put(127.0,301.0){\usebox{\plotpoint}}
\put(128.0,301.0){\usebox{\plotpoint}}
\put(128.0,302.0){\usebox{\plotpoint}}
\put(129.0,302.0){\rule[-0.200pt]{0.400pt}{0.482pt}}
\put(129.0,304.0){\usebox{\plotpoint}}
\put(130.0,304.0){\usebox{\plotpoint}}
\put(130.0,305.0){\usebox{\plotpoint}}
\put(131.0,305.0){\rule[-0.200pt]{0.400pt}{0.482pt}}
\put(131.0,307.0){\usebox{\plotpoint}}
\put(132.0,307.0){\usebox{\plotpoint}}
\put(132.0,308.0){\usebox{\plotpoint}}
\put(133.0,308.0){\usebox{\plotpoint}}
\put(133.0,309.0){\usebox{\plotpoint}}
\put(134.0,309.0){\rule[-0.200pt]{0.400pt}{0.482pt}}
\put(134.0,311.0){\usebox{\plotpoint}}
\put(135.0,311.0){\usebox{\plotpoint}}
\put(135.0,312.0){\usebox{\plotpoint}}
\put(136.0,312.0){\usebox{\plotpoint}}
\put(136.0,313.0){\usebox{\plotpoint}}
\put(137.0,313.0){\usebox{\plotpoint}}
\put(137.0,314.0){\usebox{\plotpoint}}
\put(138.0,314.0){\rule[-0.200pt]{0.400pt}{0.482pt}}
\put(138.0,316.0){\usebox{\plotpoint}}
\put(139.0,316.0){\usebox{\plotpoint}}
\put(139.0,317.0){\usebox{\plotpoint}}
\put(140.0,317.0){\usebox{\plotpoint}}
\put(140.0,318.0){\usebox{\plotpoint}}
\put(141.0,318.0){\usebox{\plotpoint}}
\put(141.0,319.0){\usebox{\plotpoint}}
\put(142.0,319.0){\usebox{\plotpoint}}
\put(142.0,320.0){\usebox{\plotpoint}}
\put(143.0,320.0){\usebox{\plotpoint}}
\put(143.0,321.0){\usebox{\plotpoint}}
\put(144.0,321.0){\usebox{\plotpoint}}
\put(144.0,322.0){\usebox{\plotpoint}}
\put(145.0,322.0){\rule[-0.200pt]{0.400pt}{0.482pt}}
\put(145.0,324.0){\usebox{\plotpoint}}
\put(146.0,324.0){\usebox{\plotpoint}}
\put(146.0,325.0){\usebox{\plotpoint}}
\put(147.0,325.0){\usebox{\plotpoint}}
\put(147.0,326.0){\usebox{\plotpoint}}
\put(148.0,326.0){\usebox{\plotpoint}}
\put(148.0,327.0){\usebox{\plotpoint}}
\put(149.0,327.0){\usebox{\plotpoint}}
\put(149.0,328.0){\usebox{\plotpoint}}
\put(150.0,328.0){\usebox{\plotpoint}}
\put(150.0,329.0){\usebox{\plotpoint}}
\put(151.0,329.0){\usebox{\plotpoint}}
\put(151.0,330.0){\usebox{\plotpoint}}
\put(152.0,330.0){\usebox{\plotpoint}}
\put(152.0,331.0){\usebox{\plotpoint}}
\put(153.0,331.0){\usebox{\plotpoint}}
\put(153.0,332.0){\usebox{\plotpoint}}
\put(154.0,332.0){\usebox{\plotpoint}}
\put(154.0,333.0){\usebox{\plotpoint}}
\put(155.0,333.0){\usebox{\plotpoint}}
\put(156,333.67){\rule{0.241pt}{0.400pt}}
\multiput(156.00,333.17)(0.500,1.000){2}{\rule{0.120pt}{0.400pt}}
\put(155.0,334.0){\usebox{\plotpoint}}
\put(157,335){\usebox{\plotpoint}}
\put(157,335){\usebox{\plotpoint}}
\put(157,335){\usebox{\plotpoint}}
\put(157,335){\usebox{\plotpoint}}
\put(157,335){\usebox{\plotpoint}}
\put(157,335){\usebox{\plotpoint}}
\put(157,335){\usebox{\plotpoint}}
\put(157,335){\usebox{\plotpoint}}
\put(157,335){\usebox{\plotpoint}}
\put(157,335){\usebox{\plotpoint}}
\put(157,335){\usebox{\plotpoint}}
\put(157,335){\usebox{\plotpoint}}
\put(157,335){\usebox{\plotpoint}}
\put(157,335){\usebox{\plotpoint}}
\put(157,335){\usebox{\plotpoint}}
\put(157,335){\usebox{\plotpoint}}
\put(157,335){\usebox{\plotpoint}}
\put(157,335){\usebox{\plotpoint}}
\put(157,335){\usebox{\plotpoint}}
\put(157,335){\usebox{\plotpoint}}
\put(157,335){\usebox{\plotpoint}}
\put(157,335){\usebox{\plotpoint}}
\put(157,335){\usebox{\plotpoint}}
\put(157,335){\usebox{\plotpoint}}
\put(157,335){\usebox{\plotpoint}}
\put(157,335){\usebox{\plotpoint}}
\put(157,335){\usebox{\plotpoint}}
\put(157,335){\usebox{\plotpoint}}
\put(157,335){\usebox{\plotpoint}}
\put(157,335){\usebox{\plotpoint}}
\put(157,335){\usebox{\plotpoint}}
\put(157,335){\usebox{\plotpoint}}
\put(157,335){\usebox{\plotpoint}}
\put(157,335){\usebox{\plotpoint}}
\put(157,335){\usebox{\plotpoint}}
\put(157,335){\usebox{\plotpoint}}
\put(157,335){\usebox{\plotpoint}}
\put(157,335){\usebox{\plotpoint}}
\put(157,335){\usebox{\plotpoint}}
\put(157,335){\usebox{\plotpoint}}
\put(157,335){\usebox{\plotpoint}}
\put(157,335){\usebox{\plotpoint}}
\put(157,335){\usebox{\plotpoint}}
\put(157,335){\usebox{\plotpoint}}
\put(157,335){\usebox{\plotpoint}}
\put(157,335){\usebox{\plotpoint}}
\put(157,335){\usebox{\plotpoint}}
\put(157,335){\usebox{\plotpoint}}
\put(157,335){\usebox{\plotpoint}}
\put(157,335){\usebox{\plotpoint}}
\put(157,335){\usebox{\plotpoint}}
\put(157,335){\usebox{\plotpoint}}
\put(157,335){\usebox{\plotpoint}}
\put(157,335){\usebox{\plotpoint}}
\put(157,335){\usebox{\plotpoint}}
\put(157,335){\usebox{\plotpoint}}
\put(157,335){\usebox{\plotpoint}}
\put(157,335){\usebox{\plotpoint}}
\put(157,335){\usebox{\plotpoint}}
\put(157,335){\usebox{\plotpoint}}
\put(157,335){\usebox{\plotpoint}}
\put(157,335){\usebox{\plotpoint}}
\put(157,335){\usebox{\plotpoint}}
\put(157,335){\usebox{\plotpoint}}
\put(157,335){\usebox{\plotpoint}}
\put(157,335){\usebox{\plotpoint}}
\put(157,335){\usebox{\plotpoint}}
\put(157,335){\usebox{\plotpoint}}
\put(157,335){\usebox{\plotpoint}}
\put(157,335){\usebox{\plotpoint}}
\put(157,335){\usebox{\plotpoint}}
\put(157,335){\usebox{\plotpoint}}
\put(157,335){\usebox{\plotpoint}}
\put(157,335){\usebox{\plotpoint}}
\put(157,335){\usebox{\plotpoint}}
\put(157,335){\usebox{\plotpoint}}
\put(157,335){\usebox{\plotpoint}}
\put(157,335){\usebox{\plotpoint}}
\put(157,335){\usebox{\plotpoint}}
\put(157,335){\usebox{\plotpoint}}
\put(157,335){\usebox{\plotpoint}}
\put(157,335){\usebox{\plotpoint}}
\put(157,335){\usebox{\plotpoint}}
\put(157,335){\usebox{\plotpoint}}
\put(157,335){\usebox{\plotpoint}}
\put(157,335){\usebox{\plotpoint}}
\put(157,335){\usebox{\plotpoint}}
\put(157,335){\usebox{\plotpoint}}
\put(157,335){\usebox{\plotpoint}}
\put(157,335){\usebox{\plotpoint}}
\put(157,335){\usebox{\plotpoint}}
\put(157,335){\usebox{\plotpoint}}
\put(157,335){\usebox{\plotpoint}}
\put(157,335){\usebox{\plotpoint}}
\put(157,335){\usebox{\plotpoint}}
\put(157,335){\usebox{\plotpoint}}
\put(157,335){\usebox{\plotpoint}}
\put(157,335){\usebox{\plotpoint}}
\put(157,335){\usebox{\plotpoint}}
\put(157,335){\usebox{\plotpoint}}
\put(157,335){\usebox{\plotpoint}}
\put(157,335){\usebox{\plotpoint}}
\put(157,335){\usebox{\plotpoint}}
\put(157,335){\usebox{\plotpoint}}
\put(157,335){\usebox{\plotpoint}}
\put(157,335){\usebox{\plotpoint}}
\put(157,335){\usebox{\plotpoint}}
\put(157,335){\usebox{\plotpoint}}
\put(157,335){\usebox{\plotpoint}}
\put(157,335){\usebox{\plotpoint}}
\put(157,335){\usebox{\plotpoint}}
\put(157.0,335.0){\usebox{\plotpoint}}
\put(158.0,335.0){\usebox{\plotpoint}}
\put(158.0,336.0){\usebox{\plotpoint}}
\put(159.0,336.0){\usebox{\plotpoint}}
\put(159.0,337.0){\usebox{\plotpoint}}
\put(160.0,337.0){\usebox{\plotpoint}}
\put(160.0,338.0){\usebox{\plotpoint}}
\put(161.0,338.0){\usebox{\plotpoint}}
\put(161.0,339.0){\usebox{\plotpoint}}
\put(162.0,339.0){\usebox{\plotpoint}}
\put(162.0,340.0){\usebox{\plotpoint}}
\put(163.0,340.0){\usebox{\plotpoint}}
\put(163.0,341.0){\rule[-0.200pt]{0.482pt}{0.400pt}}
\put(165.0,341.0){\usebox{\plotpoint}}
\put(165.0,342.0){\usebox{\plotpoint}}
\put(166.0,342.0){\usebox{\plotpoint}}
\put(166.0,343.0){\usebox{\plotpoint}}
\put(167.0,343.0){\usebox{\plotpoint}}
\put(167.0,344.0){\usebox{\plotpoint}}
\put(168.0,344.0){\usebox{\plotpoint}}
\put(168.0,345.0){\rule[-0.200pt]{0.482pt}{0.400pt}}
\put(170.0,345.0){\usebox{\plotpoint}}
\put(170.0,346.0){\usebox{\plotpoint}}
\put(171.0,346.0){\usebox{\plotpoint}}
\put(171.0,347.0){\usebox{\plotpoint}}
\put(172.0,347.0){\usebox{\plotpoint}}
\put(172.0,348.0){\rule[-0.200pt]{0.482pt}{0.400pt}}
\put(174.0,348.0){\usebox{\plotpoint}}
\put(174.0,349.0){\usebox{\plotpoint}}
\put(175.0,349.0){\usebox{\plotpoint}}
\put(175.0,350.0){\rule[-0.200pt]{0.482pt}{0.400pt}}
\put(177.0,350.0){\usebox{\plotpoint}}
\put(177.0,351.0){\usebox{\plotpoint}}
\put(178.0,351.0){\usebox{\plotpoint}}
\put(178.0,352.0){\usebox{\plotpoint}}
\put(179.0,352.0){\usebox{\plotpoint}}
\put(179.0,353.0){\rule[-0.200pt]{0.482pt}{0.400pt}}
\put(181.0,353.0){\usebox{\plotpoint}}
\put(181.0,354.0){\usebox{\plotpoint}}
\put(182.0,354.0){\usebox{\plotpoint}}
\put(182.0,355.0){\rule[-0.200pt]{0.482pt}{0.400pt}}
\put(184.0,355.0){\usebox{\plotpoint}}
\put(184.0,356.0){\rule[-0.200pt]{0.482pt}{0.400pt}}
\put(186.0,356.0){\usebox{\plotpoint}}
\put(186.0,357.0){\usebox{\plotpoint}}
\put(187.0,357.0){\usebox{\plotpoint}}
\put(187.0,358.0){\rule[-0.200pt]{0.482pt}{0.400pt}}
\put(189.0,358.0){\usebox{\plotpoint}}
\put(189.0,359.0){\rule[-0.200pt]{0.482pt}{0.400pt}}
\put(191.0,359.0){\usebox{\plotpoint}}
\put(191.0,360.0){\usebox{\plotpoint}}
\put(192.0,360.0){\usebox{\plotpoint}}
\put(192.0,361.0){\rule[-0.200pt]{0.482pt}{0.400pt}}
\put(194.0,361.0){\usebox{\plotpoint}}
\put(194.0,362.0){\rule[-0.200pt]{0.482pt}{0.400pt}}
\put(196.0,362.0){\usebox{\plotpoint}}
\put(196.0,363.0){\rule[-0.200pt]{0.482pt}{0.400pt}}
\put(198.0,363.0){\usebox{\plotpoint}}
\put(198.0,364.0){\usebox{\plotpoint}}
\put(199.0,364.0){\usebox{\plotpoint}}
\put(199.0,365.0){\rule[-0.200pt]{0.482pt}{0.400pt}}
\put(201.0,365.0){\usebox{\plotpoint}}
\put(201.0,366.0){\rule[-0.200pt]{0.482pt}{0.400pt}}
\put(203.0,366.0){\usebox{\plotpoint}}
\put(203.0,367.0){\rule[-0.200pt]{0.482pt}{0.400pt}}
\put(205.0,367.0){\usebox{\plotpoint}}
\put(205.0,368.0){\rule[-0.200pt]{0.482pt}{0.400pt}}
\put(207.0,368.0){\usebox{\plotpoint}}
\put(207.0,369.0){\rule[-0.200pt]{0.482pt}{0.400pt}}
\put(209.0,369.0){\usebox{\plotpoint}}
\put(209.0,370.0){\rule[-0.200pt]{0.482pt}{0.400pt}}
\put(211.0,370.0){\usebox{\plotpoint}}
\put(211.0,371.0){\rule[-0.200pt]{0.723pt}{0.400pt}}
\put(214.0,371.0){\usebox{\plotpoint}}
\put(214.0,372.0){\rule[-0.200pt]{0.482pt}{0.400pt}}
\put(216.0,372.0){\usebox{\plotpoint}}
\put(216.0,373.0){\rule[-0.200pt]{0.482pt}{0.400pt}}
\put(218.0,373.0){\usebox{\plotpoint}}
\put(218.0,374.0){\rule[-0.200pt]{0.482pt}{0.400pt}}
\put(220.0,374.0){\usebox{\plotpoint}}
\put(220.0,375.0){\rule[-0.200pt]{0.723pt}{0.400pt}}
\put(223.0,375.0){\usebox{\plotpoint}}
\put(223.0,376.0){\rule[-0.200pt]{0.482pt}{0.400pt}}
\put(225.0,376.0){\usebox{\plotpoint}}
\put(225.0,377.0){\rule[-0.200pt]{0.723pt}{0.400pt}}
\put(228.0,377.0){\usebox{\plotpoint}}
\put(228.0,378.0){\rule[-0.200pt]{0.482pt}{0.400pt}}
\put(230.0,378.0){\usebox{\plotpoint}}
\put(230.0,379.0){\rule[-0.200pt]{0.723pt}{0.400pt}}
\put(233.0,379.0){\usebox{\plotpoint}}
\put(235,379.67){\rule{0.241pt}{0.400pt}}
\multiput(235.00,379.17)(0.500,1.000){2}{\rule{0.120pt}{0.400pt}}
\put(233.0,380.0){\rule[-0.200pt]{0.482pt}{0.400pt}}
\put(236,381){\usebox{\plotpoint}}
\put(236,381){\usebox{\plotpoint}}
\put(236,381){\usebox{\plotpoint}}
\put(236,381){\usebox{\plotpoint}}
\put(236,381){\usebox{\plotpoint}}
\put(236,381){\usebox{\plotpoint}}
\put(236,381){\usebox{\plotpoint}}
\put(236,381){\usebox{\plotpoint}}
\put(236,381){\usebox{\plotpoint}}
\put(236,381){\usebox{\plotpoint}}
\put(236,381){\usebox{\plotpoint}}
\put(236,381){\usebox{\plotpoint}}
\put(236,381){\usebox{\plotpoint}}
\put(236,381){\usebox{\plotpoint}}
\put(236,381){\usebox{\plotpoint}}
\put(236,381){\usebox{\plotpoint}}
\put(236,381){\usebox{\plotpoint}}
\put(236,381){\usebox{\plotpoint}}
\put(236,381){\usebox{\plotpoint}}
\put(236,381){\usebox{\plotpoint}}
\put(236,381){\usebox{\plotpoint}}
\put(236,381){\usebox{\plotpoint}}
\put(236,381){\usebox{\plotpoint}}
\put(236,381){\usebox{\plotpoint}}
\put(236,381){\usebox{\plotpoint}}
\put(236,381){\usebox{\plotpoint}}
\put(236,381){\usebox{\plotpoint}}
\put(236,381){\usebox{\plotpoint}}
\put(236,381){\usebox{\plotpoint}}
\put(236,381){\usebox{\plotpoint}}
\put(236,381){\usebox{\plotpoint}}
\put(236,381){\usebox{\plotpoint}}
\put(236,381){\usebox{\plotpoint}}
\put(236,381){\usebox{\plotpoint}}
\put(236,381){\usebox{\plotpoint}}
\put(236,381){\usebox{\plotpoint}}
\put(236,381){\usebox{\plotpoint}}
\put(236,381){\usebox{\plotpoint}}
\put(236,381){\usebox{\plotpoint}}
\put(236,381){\usebox{\plotpoint}}
\put(236,381){\usebox{\plotpoint}}
\put(236,381){\usebox{\plotpoint}}
\put(236,381){\usebox{\plotpoint}}
\put(236,381){\usebox{\plotpoint}}
\put(236,381){\usebox{\plotpoint}}
\put(236,381){\usebox{\plotpoint}}
\put(236,381){\usebox{\plotpoint}}
\put(236,381){\usebox{\plotpoint}}
\put(236,381){\usebox{\plotpoint}}
\put(236,381){\usebox{\plotpoint}}
\put(236,381){\usebox{\plotpoint}}
\put(236,381){\usebox{\plotpoint}}
\put(236,381){\usebox{\plotpoint}}
\put(236,381){\usebox{\plotpoint}}
\put(236,381){\usebox{\plotpoint}}
\put(236,381){\usebox{\plotpoint}}
\put(236,381){\usebox{\plotpoint}}
\put(236,381){\usebox{\plotpoint}}
\put(236,381){\usebox{\plotpoint}}
\put(236,381){\usebox{\plotpoint}}
\put(236,381){\usebox{\plotpoint}}
\put(236,381){\usebox{\plotpoint}}
\put(236,381){\usebox{\plotpoint}}
\put(236,381){\usebox{\plotpoint}}
\put(236,381){\usebox{\plotpoint}}
\put(236,381){\usebox{\plotpoint}}
\put(236,381){\usebox{\plotpoint}}
\put(236,381){\usebox{\plotpoint}}
\put(236,381){\usebox{\plotpoint}}
\put(236,381){\usebox{\plotpoint}}
\put(236,381){\usebox{\plotpoint}}
\put(236,381){\usebox{\plotpoint}}
\put(236,381){\usebox{\plotpoint}}
\put(236,381){\usebox{\plotpoint}}
\put(236,381){\usebox{\plotpoint}}
\put(236,381){\usebox{\plotpoint}}
\put(236,381){\usebox{\plotpoint}}
\put(236,381){\usebox{\plotpoint}}
\put(236,381){\usebox{\plotpoint}}
\put(236,381){\usebox{\plotpoint}}
\put(236,381){\usebox{\plotpoint}}
\put(236,381){\usebox{\plotpoint}}
\put(236,381){\usebox{\plotpoint}}
\put(236,381){\usebox{\plotpoint}}
\put(236,381){\usebox{\plotpoint}}
\put(236,381){\usebox{\plotpoint}}
\put(236,381){\usebox{\plotpoint}}
\put(236,381){\usebox{\plotpoint}}
\put(236,381){\usebox{\plotpoint}}
\put(236,381){\usebox{\plotpoint}}
\put(236,381){\usebox{\plotpoint}}
\put(236,381){\usebox{\plotpoint}}
\put(236,381){\usebox{\plotpoint}}
\put(236,381){\usebox{\plotpoint}}
\put(236,381){\usebox{\plotpoint}}
\put(236,381){\usebox{\plotpoint}}
\put(236,381){\usebox{\plotpoint}}
\put(236,381){\usebox{\plotpoint}}
\put(236,381){\usebox{\plotpoint}}
\put(236,381){\usebox{\plotpoint}}
\put(236,381){\usebox{\plotpoint}}
\put(236,381){\usebox{\plotpoint}}
\put(236,381){\usebox{\plotpoint}}
\put(236,381){\usebox{\plotpoint}}
\put(236,381){\usebox{\plotpoint}}
\put(236,381){\usebox{\plotpoint}}
\put(236,381){\usebox{\plotpoint}}
\put(236,381){\usebox{\plotpoint}}
\put(236,381){\usebox{\plotpoint}}
\put(236,381){\usebox{\plotpoint}}
\put(236,381){\usebox{\plotpoint}}
\put(236.0,381.0){\rule[-0.200pt]{0.482pt}{0.400pt}}
\put(238.0,381.0){\usebox{\plotpoint}}
\put(238.0,382.0){\rule[-0.200pt]{0.723pt}{0.400pt}}
\put(241.0,382.0){\usebox{\plotpoint}}
\put(241.0,383.0){\rule[-0.200pt]{0.723pt}{0.400pt}}
\put(244.0,383.0){\usebox{\plotpoint}}
\put(244.0,384.0){\rule[-0.200pt]{0.723pt}{0.400pt}}
\put(247.0,384.0){\usebox{\plotpoint}}
\put(247.0,385.0){\rule[-0.200pt]{0.723pt}{0.400pt}}
\put(250.0,385.0){\usebox{\plotpoint}}
\put(250.0,386.0){\rule[-0.200pt]{0.723pt}{0.400pt}}
\put(253.0,386.0){\usebox{\plotpoint}}
\put(253.0,387.0){\rule[-0.200pt]{0.723pt}{0.400pt}}
\put(256.0,387.0){\usebox{\plotpoint}}
\put(256.0,388.0){\rule[-0.200pt]{0.723pt}{0.400pt}}
\put(259.0,388.0){\usebox{\plotpoint}}
\put(259.0,389.0){\rule[-0.200pt]{0.964pt}{0.400pt}}
\put(263.0,389.0){\usebox{\plotpoint}}
\put(263.0,390.0){\rule[-0.200pt]{0.723pt}{0.400pt}}
\put(266.0,390.0){\usebox{\plotpoint}}
\put(266.0,391.0){\rule[-0.200pt]{0.964pt}{0.400pt}}
\put(270.0,391.0){\usebox{\plotpoint}}
\put(270.0,392.0){\rule[-0.200pt]{0.964pt}{0.400pt}}
\put(274.0,392.0){\usebox{\plotpoint}}
\put(274.0,393.0){\rule[-0.200pt]{0.723pt}{0.400pt}}
\put(277.0,393.0){\usebox{\plotpoint}}
\put(277.0,394.0){\rule[-0.200pt]{0.964pt}{0.400pt}}
\put(281.0,394.0){\usebox{\plotpoint}}
\put(281.0,395.0){\rule[-0.200pt]{0.964pt}{0.400pt}}
\put(285.0,395.0){\usebox{\plotpoint}}
\put(285.0,396.0){\rule[-0.200pt]{0.964pt}{0.400pt}}
\put(289.0,396.0){\usebox{\plotpoint}}
\put(289.0,397.0){\rule[-0.200pt]{1.204pt}{0.400pt}}
\put(294.0,397.0){\usebox{\plotpoint}}
\put(294.0,398.0){\rule[-0.200pt]{0.964pt}{0.400pt}}
\put(298.0,398.0){\usebox{\plotpoint}}
\put(298.0,399.0){\rule[-0.200pt]{1.204pt}{0.400pt}}
\put(303.0,399.0){\usebox{\plotpoint}}
\put(303.0,400.0){\rule[-0.200pt]{0.964pt}{0.400pt}}
\put(307.0,400.0){\usebox{\plotpoint}}
\put(307.0,401.0){\rule[-0.200pt]{1.204pt}{0.400pt}}
\put(312.0,401.0){\usebox{\plotpoint}}
\put(312.0,402.0){\rule[-0.200pt]{1.204pt}{0.400pt}}
\put(317.0,402.0){\usebox{\plotpoint}}
\put(317.0,403.0){\rule[-0.200pt]{1.204pt}{0.400pt}}
\put(322.0,403.0){\usebox{\plotpoint}}
\put(322.0,404.0){\rule[-0.200pt]{1.445pt}{0.400pt}}
\put(328.0,404.0){\usebox{\plotpoint}}
\put(328.0,405.0){\rule[-0.200pt]{1.204pt}{0.400pt}}
\put(333.0,405.0){\usebox{\plotpoint}}
\put(333.0,406.0){\rule[-0.200pt]{1.445pt}{0.400pt}}
\put(339.0,406.0){\usebox{\plotpoint}}
\put(339.0,407.0){\rule[-0.200pt]{1.445pt}{0.400pt}}
\put(345.0,407.0){\usebox{\plotpoint}}
\put(345.0,408.0){\rule[-0.200pt]{1.445pt}{0.400pt}}
\put(351.0,408.0){\usebox{\plotpoint}}
\put(351.0,409.0){\rule[-0.200pt]{1.686pt}{0.400pt}}
\put(358.0,409.0){\usebox{\plotpoint}}
\put(358.0,410.0){\rule[-0.200pt]{1.686pt}{0.400pt}}
\put(365.0,410.0){\usebox{\plotpoint}}
\put(365.0,411.0){\rule[-0.200pt]{1.686pt}{0.400pt}}
\put(372.0,411.0){\usebox{\plotpoint}}
\put(372.0,412.0){\rule[-0.200pt]{1.686pt}{0.400pt}}
\put(379.0,412.0){\usebox{\plotpoint}}
\put(379.0,413.0){\rule[-0.200pt]{1.686pt}{0.400pt}}
\put(386.0,413.0){\usebox{\plotpoint}}
\put(386.0,414.0){\rule[-0.200pt]{1.927pt}{0.400pt}}
\put(394.0,414.0){\usebox{\plotpoint}}
\put(394.0,415.0){\rule[-0.200pt]{2.168pt}{0.400pt}}
\put(403.0,415.0){\usebox{\plotpoint}}
\put(403.0,416.0){\rule[-0.200pt]{1.927pt}{0.400pt}}
\put(411.0,416.0){\usebox{\plotpoint}}
\put(411.0,417.0){\rule[-0.200pt]{2.168pt}{0.400pt}}
\put(420.0,417.0){\usebox{\plotpoint}}
\put(420.0,418.0){\rule[-0.200pt]{2.409pt}{0.400pt}}
\put(430.0,418.0){\usebox{\plotpoint}}
\put(430.0,419.0){\rule[-0.200pt]{2.409pt}{0.400pt}}
\put(440.0,419.0){\usebox{\plotpoint}}
\put(440.0,420.0){\rule[-0.200pt]{2.409pt}{0.400pt}}
\put(450.0,420.0){\usebox{\plotpoint}}
\put(450.0,421.0){\rule[-0.200pt]{2.650pt}{0.400pt}}
\put(461.0,421.0){\usebox{\plotpoint}}
\put(461.0,422.0){\rule[-0.200pt]{2.650pt}{0.400pt}}
\put(472.0,422.0){\usebox{\plotpoint}}
\put(472.0,423.0){\rule[-0.200pt]{3.132pt}{0.400pt}}
\put(485.0,423.0){\usebox{\plotpoint}}
\put(485.0,424.0){\rule[-0.200pt]{2.891pt}{0.400pt}}
\put(497.0,424.0){\usebox{\plotpoint}}
\put(497.0,425.0){\rule[-0.200pt]{3.373pt}{0.400pt}}
\put(511.0,425.0){\usebox{\plotpoint}}
\put(511.0,426.0){\rule[-0.200pt]{3.373pt}{0.400pt}}
\put(525.0,426.0){\usebox{\plotpoint}}
\put(525.0,427.0){\rule[-0.200pt]{3.613pt}{0.400pt}}
\put(540.0,427.0){\usebox{\plotpoint}}
\put(540.0,428.0){\rule[-0.200pt]{3.854pt}{0.400pt}}
\put(556.0,428.0){\usebox{\plotpoint}}
\put(556.0,429.0){\rule[-0.200pt]{4.095pt}{0.400pt}}
\put(573.0,429.0){\usebox{\plotpoint}}
\put(573.0,430.0){\rule[-0.200pt]{4.577pt}{0.400pt}}
\put(592.0,430.0){\usebox{\plotpoint}}
\put(592.0,431.0){\rule[-0.200pt]{4.577pt}{0.400pt}}
\put(611.0,431.0){\usebox{\plotpoint}}
\put(611.0,432.0){\rule[-0.200pt]{5.059pt}{0.400pt}}
\put(632.0,432.0){\usebox{\plotpoint}}
\put(632.0,433.0){\rule[-0.200pt]{5.300pt}{0.400pt}}
\put(654.0,433.0){\usebox{\plotpoint}}
\put(654.0,434.0){\rule[-0.200pt]{5.782pt}{0.400pt}}
\put(678.0,434.0){\usebox{\plotpoint}}
\put(678.0,435.0){\rule[-0.200pt]{6.263pt}{0.400pt}}
\put(704.0,435.0){\usebox{\plotpoint}}
\put(704.0,436.0){\rule[-0.200pt]{6.745pt}{0.400pt}}
\put(732.0,436.0){\usebox{\plotpoint}}
\put(732.0,437.0){\rule[-0.200pt]{7.227pt}{0.400pt}}
\put(762.0,437.0){\usebox{\plotpoint}}
\put(762.0,438.0){\rule[-0.200pt]{7.950pt}{0.400pt}}
\put(795.0,438.0){\usebox{\plotpoint}}
\put(795.0,439.0){\rule[-0.200pt]{8.913pt}{0.400pt}}
\put(832.0,439.0){\usebox{\plotpoint}}
\put(832.0,440.0){\rule[-0.200pt]{9.395pt}{0.400pt}}
\put(871.0,440.0){\usebox{\plotpoint}}
\put(871.0,441.0){\rule[-0.200pt]{10.600pt}{0.400pt}}
\put(915.0,441.0){\usebox{\plotpoint}}
\put(915.0,442.0){\rule[-0.200pt]{9.154pt}{0.400pt}}
\put(60,448){\usebox{\plotpoint}}
\put(60.00,448.00){\usebox{\plotpoint}}
\put(80.76,448.00){\usebox{\plotpoint}}
\put(101.51,448.00){\usebox{\plotpoint}}
\put(122.27,448.00){\usebox{\plotpoint}}
\put(143.02,448.00){\usebox{\plotpoint}}
\put(163.78,448.00){\usebox{\plotpoint}}
\put(184.53,448.00){\usebox{\plotpoint}}
\put(205.29,448.00){\usebox{\plotpoint}}
\put(226.04,448.00){\usebox{\plotpoint}}
\put(246.80,448.00){\usebox{\plotpoint}}
\put(267.56,448.00){\usebox{\plotpoint}}
\put(288.31,448.00){\usebox{\plotpoint}}
\put(309.07,448.00){\usebox{\plotpoint}}
\put(329.82,448.00){\usebox{\plotpoint}}
\put(350.58,448.00){\usebox{\plotpoint}}
\put(371.33,448.00){\usebox{\plotpoint}}
\put(392.09,448.00){\usebox{\plotpoint}}
\put(412.84,448.00){\usebox{\plotpoint}}
\put(433.60,448.00){\usebox{\plotpoint}}
\put(454.35,448.00){\usebox{\plotpoint}}
\put(475.11,448.00){\usebox{\plotpoint}}
\put(495.87,448.00){\usebox{\plotpoint}}
\put(516.62,448.00){\usebox{\plotpoint}}
\put(537.38,448.00){\usebox{\plotpoint}}
\put(558.13,448.00){\usebox{\plotpoint}}
\put(578.89,448.00){\usebox{\plotpoint}}
\put(599.64,448.00){\usebox{\plotpoint}}
\put(620.40,448.00){\usebox{\plotpoint}}
\put(641.15,448.00){\usebox{\plotpoint}}
\put(661.91,448.00){\usebox{\plotpoint}}
\put(682.66,448.00){\usebox{\plotpoint}}
\put(703.42,448.00){\usebox{\plotpoint}}
\put(724.18,448.00){\usebox{\plotpoint}}
\put(744.93,448.00){\usebox{\plotpoint}}
\put(765.69,448.00){\usebox{\plotpoint}}
\put(786.44,448.00){\usebox{\plotpoint}}
\put(807.20,448.00){\usebox{\plotpoint}}
\put(827.95,448.00){\usebox{\plotpoint}}
\put(848.71,448.00){\usebox{\plotpoint}}
\put(869.46,448.00){\usebox{\plotpoint}}
\put(890.22,448.00){\usebox{\plotpoint}}
\put(910.98,448.00){\usebox{\plotpoint}}
\put(931.73,448.00){\usebox{\plotpoint}}
\put(952.49,448.00){\usebox{\plotpoint}}
\put(953,448){\usebox{\plotpoint}}
\end{picture}
\end{center}
\caption{critical temperatures ratio versus coupling constants ratio $\lambda/4e^2$}
\end{figure}

The intriguing feature is that both the two superconducting
phases could be revealed in a single system, the effective
measurability depending on the ratio $\lambda /4e^2$ or,
ultimately, on the particular medium considered. In fact, since the energy
density scales as the fourth power of the temperature, it becomes
energetically favorable for the system to pass from the phase with
the higher critical temperature ($T_1$) to that with a lower one
($T_2$).

Let we stress that when considering 
the two limiting cases $\lambda\gg 4e^2$ and $\lambda\ll4e^2$, 
we do not refer to the transition between a
strong-coupling BCS to a weak-coupling one. The two Cooper pair
condensates we consider (described by the real order parameters
$\phi_1$, $\phi_2$ or $\eta$, $\theta$ respectively) differ, indeed, 
by their self-interaction with respect to the electromagnetic interaction:
but, in any case, such a difference is quite small, and no
transition from one regime (strong-coupling BCS) to the other
(weak-coupling BCS) occurs. This may be also deduced from the fact
that, to the best, the change in the critical temperature is at
most of 15\%.

In principle we are allowed to take the only quadratic terms in the field disgregarding higher order terms (this is truly correct only near a critical point), since the two critical temperatures 
do not differ very much. However we have not made such an approximation:
In our model there are two different critical temperatures, but the present
quadratic approximation applies to two different fields representations, not overlapping, and a given gauge holds only near one of the two critical points.
Near a critical temperature we study the system by means of two
real fields whilst, near the other temperature, two different degrees
of freedom describe correctly the same system. 
As above said, only for a infinite density medium, \ $\eta_0\to\infty$, \ 
such order parameters are physically equivalent.

Summing up, we can envisage the following scenario. By lowering
the temperature in a GL superconductor, it undergoes a first phase
transition at $T =T_1$; electrons con condensate into Cooper pairs
described by the field $\eta$, while the degrees of freedom
related to $\theta$ are ``absorbed'' in a massive gauge field
$A_\mu$ describing finite range electromagnetic interactions. With a further
lowering of the temperature, different (but not additional)
degrees of the freedom of the condensate are excited, described by
the fields $\phi_1 ,\phi_2$, and another second-order phase
transition takes place at $T=T_2< T_1$. This phase also exhibits
superconducting state features, but the direct role of the phase
field $\theta$ played in the excitations of $\phi_1, \phi_2$ [see
Eqs.\,(\ref{8b})] may reveal the possible appearance of vortices.
As is well known, in fact, such topological objects arise when the net variation
of the local phase field $\theta(x)$ on a given closed loop $\gamma$ 
does not vanish: 
\begin{equation}
\oint_\gamma \vecna \theta \cdot {\rm d}{\ellbf} = 2\pi\,;
\label{phase}
\end{equation}
this, being a typical quantum topological effect, should be mostly detectable
at very low temperatures.
Although the long-range order in the superconductors, described by the complex 
scalar field $\phi$, is not affected by a small concentration of these topological defects, 
this is no longer true for larger defects.    
The net effect of the vortices is that of a weakening of the finite range 
electromagnetic interaction between electrons with respect to 
the Cooper pair self-interaction, with a lowering of the critical temperature:
\begin{equation}
T_1 = \sqrt{\frac{4m^2}{\lambda + 4e^2}} \ \longrightarrow \ T_2 =
\sqrt{\frac{4m^2}{\displaystyle{\frac{4}{3}}\lambda + 4e^2}}. \label{12}
\end{equation}
Though the detection of different properties in the two
superconducting phases may be quite involved, nevertheless the
effect could be as large as $15\%$ depending on $\lambda/4e^2$, so
that the experimental search in peculiar materials should be
highly stimulated. \\
Similar effects might be expected also in superfluid Helium phenomena, 
where the Higgs mechanism envisaged above should apply as well, so that 
the experimental search for them may proceed in this framework too.\\
Moreover, being the present approach very general, we might think at a possible
two-phases behavior also in broken gauge symmetries of fundamental forces of Nature, 
so that the peculiar mechanism here envisaged might have 
implications in the Higgs phenomena occurring in the early Universe.

\vspace{1cm}

\noindent {\large{\bf Acknowledgments}}

\vspace*{0.2cm}

\noindent We are grateful to an anonymous Referee for useful suggestions,
and to V. Cataudella, G. De Filippis, V. Marigliano 
Ramaglia and A. Naddeo for very stimulating discussions about some topics
here covered.

\

\

\end{document}